\documentclass[12pt,preprint]{aastex}

\begin{document}

\title{Deep Spitzer spectroscopy of the `Flying Saucer' edge-on disk: Large
grains beyond 50 AU.}

\author{Klaus M. Pontoppidan\altaffilmark{1,2}}
\altaffiltext{1}{California Institute of Technology, Division of Geological and Planetary Sciences, MS 150-21, Pasadena, CA 91125}
\altaffiltext{2}{Hubble Fellow}
\email{pontoppi@gps.caltech.edu}

\author{Karl R. Stapelfeldt\altaffilmark{3}}
\altaffiltext{3}{Jet Propulsion Laboratory, 4800 Oak Grove Drive, Pasadena, CA 91109} 
\email{krs@exoplanet.jpl.nasa.gov}

\author{Geoffrey A. Blake\altaffilmark{1}}
\email{gab@gps.caltech.edu}

\author{Ewine F. van Dishoeck\altaffilmark{4}}
\altaffiltext{4}{Leiden Observatory, P.O.Box 9513, NL-2300 RA Leiden, The Netherlands}
\email{ewine@strw.leidenuniv.nl}

\author{Cornelis P. Dullemond\altaffilmark{5}}
\altaffiltext{5}{Max-Planck-Institut f{\"u}r Astronomie, K{\"o}nigstuhl 17, D69117 Heidelberg, Germany}
\email{dullemon@mpia-hd.mpg.de}

\begin{abstract}
We present deep Spitzer-IRS low-resolution ($\lambda/\Delta\lambda\sim$100) 5-35\,$\mu$m 
spectroscopy of the edge-on disk
``the Flying Saucer'' (2MASS J16281370-2431391) in the Ophiuchus molecular cloud. The spectral
energy distribution exhibits the
characteristic two-peak shape predicted for a circumstellar disk viewed very close to 
edge-on. The short-wavelength peak is entirely due to photons scattered off
the surface of the disk, while the long-wavelength peak is due to thermal
emission from the disk itself. The Spitzer spectrum represents the first spectroscopic detection of
scattered light out to 15\,$\mu$m from a bona-fide, isolated edge-on disk around a 
T Tauri star. The depth and the wavelength of the mid-infrared "valley" of the SED give direct constraints on the
size distribution of large grains in the disk. Using a 2D continuum radiative transfer model, we find that a
significant amount of 5-10 $\mu$m-sized grains is required in the surface layers of the disk at radii
of 50-300\,AU. 
The detection of relatively large grains in the upper layers 
implies that vertical mixing is effective, since grain growth models predict the grains would otherwise settle
deep in the disk on short time scales.
Additionally, we {\it tentatively} detect the 9.66\,$\mu$m S(3) line of H$_2$ and the 
11.2\,$\mu$m emission feature due to PAHs. 

\end{abstract}


\section{Introduction}
Highly inclined circumstellar disks are particularly well-suited for 
studies of the dust properties in optically thick proto-planetary systems. The very
high degree of obscuration toward the bright inner parts of these disks
acts as a natural coronograph and thereby increases the contrast to the fainter outer parts by several orders 
of magnitude, enabling direct studies of the optical and infrared properties of the disk surface layers at
radii of $\gtrsim$50\,AU; the region of the $\tau=1$ surface to scattering as seen by an observer. In particular, the scattering properties and therefore the grain
size distributions and the vertical height of the disk can be constrained using spatially resolved imaging. Examples of such studies employing
optical and near-infrared scattered light images of edge-on disks include
\cite{Burrows96,Stapelfeldt98,Cotera01,Stapelfeldt03,Wolf03}. 

These studies have shown that the extended emission at optical and near-infrared wavelengths has a morphology
closely resembling that theoretically expected for edge-on disks, i.e. two flattened reflection nebulosities bisected by
a dark lane in the disk plane. 
Attempts to use these relatively short wavelengths to search for grains larger than those found in the interstellar medium have 
produced somewhat inconclusive results, and it has become clear that observations at longer wavelengths are required \citep[e.g.][]{Brandner00,Wood02}. 
A number of mid-infrared images have been made of edge-on disks.
Spatially resolved emission at 11.8\,$\mu$m from the edge-on disk
HK Tau B was presented by \cite{McCabe03}, who used the image to infer a grain size distribution
with relatively large grains (1.5--3.0\,$\mu$m). \cite{Perrin06} presented mid-infrared 
images of an edge-on disk clearly dominated by Polycyclic Aromatic Hydrocarbon (PAH) 
emission rather than scattered light.

A problem is that edge-on disks tend to be very faint in the mid-infrared, requiring
very sensitive observational facilities. Highly sensitive spectroscopy offered by the {\it Spitzer Space Telescope} provides new possibilities. 
Spitzer will not spatially resolve the mid-infrared emission, except for extremely large disks, but can obtain spectroscopic information
that provides essential constraints for systems that have already been imaged at shorter wavelengths.  
One important advantage of spectroscopy is that line emission can be
distinguished from scattered light, a distinction that 
becomes essential when trying to constrain scattering properties of the dust. 

In this letter, we report deep mid-infrared spectroscopic observations using the Infrared Spectrograph
(IRS) on Spitzer of the 
``Flying Saucer'' edge-on disk located on the outskirts of the Ophiuchus star forming cloud core,
first identified in high resolution near-infrared images \citep{Grosso03}. The spectrum
was obtained as part
of the {\it ``From cores to disks'' } (c2d)
Legacy program \citep{Evans03}. The ``Flying Saucer'' is one of the most
isolated edge-on disks known with a foreground visual 
extinction less than a few magnitudes, and is viewed almost exactly edge-on 
($i\gtrsim 85\degr$).

\section{Data reduction and results}

\begin{figure}
  \plotone{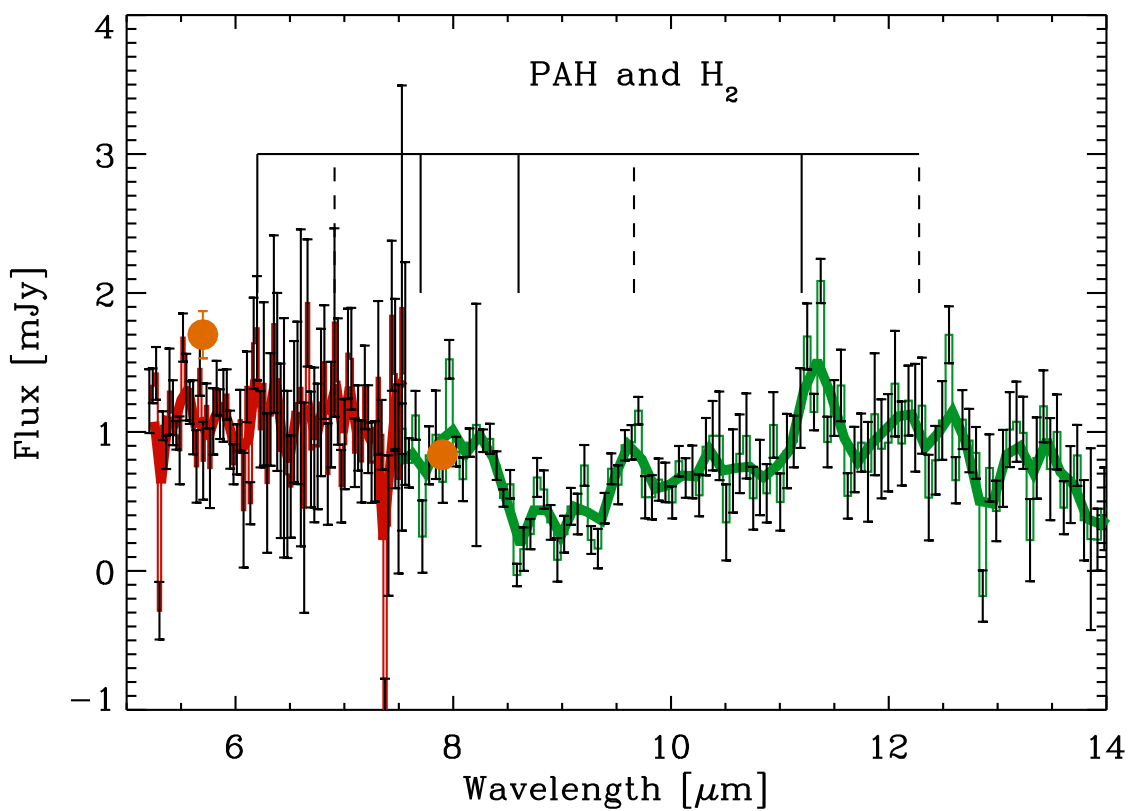}
  \caption{The short-low spectrum showing the tentative emission features at 9.7 and 11.2\,$\mu$m. The vertical bars show the location
of common PAH features (solid bars) as well as the S(2), S(3) and S(4) lines of molecular hydrogen (dashed bars). The filled circles show the 
IRAC photometry in bands 3 and 4. The thick curve shows a spectrum that has been smoothed with a two-point box car filter.}
  \label{saucer_pah}
\end{figure}

The ``Flying Saucer'' (2MASS J16281370-2431391, 
$\alpha=$16$^h$28$^m$13\fs7, $\delta=$-24\degr31\arcmin39\arcsec [J2000]) 
was observed using both the short-low (SL2 = 5.2--7.5\,$\mu$m, SL1 = 7.5--14.5\,$\mu$m) and long-low (LL2 = 14.2--21.0\,$\mu$m, LL1 = 19.8--38.1\,$\mu$m) modules
of the IRS (AOR KEY 0009829632). 
Each SL order was observed for $10\times 61\,$s and
each LL order for $20\times 31.5$s. The spectral extraction was performed
from the BCD pipeline (version S13.2.0) products by co-adding each nod position and 
then subtracting the two resulting 2D spectra to effectively remove the
background. Due to the presence of a bright source on the peakup array for the SL1 observation, 
the noise at wavelengths between 12.7 and 14.5\,$\mu$m is significantly higher than in 
other parts of the spectrum. The ``Flying Saucer'' is clearly detected in SL1 and 2 as well as in LL1; there is no detection
in LL2, consistent with a location of the flux minimum around 20\,$\mu$m. One-dimensional spectra
were extracted using 3-pixel (5.4\arcsec) apertures to optimize the signal-to-noise. 
To aid the modeling efforts, IRAC and MIPS photometry has been extracted from the 
c2d photometric database. The spectra match 
the IRAC band 3 (5.7\,$\mu$m) and 4 (7.9\,$\mu$m) photometry within the accuracy of the absolute calibration. The difference in
slope between the spectrum and IRAC 3 and 4 may be due to the source being extended; the SL slit width is 3.6\arcsec. 
The source is detected in MIPS1 (24\,$\mu$m) and only an upper limit is available
in MIPS2 (70\,$\mu$m) (see Table \ref{photometry}).

The resulting SL spectrum is shown in Fig. \ref{saucer_pah} and shows two weak, narrow emission features at $\sim 9.7\,\mu$m and $\sim 11.2\,\mu$m with 
line-to-continuum ratios of $\sim 1$. Care should be taken in interpreting these features because the
behaviour of the spectrograph is not yet fully understood, and features at the 5-10\% level may be spurious. 
Assuming gaussian noise, the two emission features are detected with S/N ratios of 4-5. Inspection of the co-added 2D spectra
confirms that the features are compact and centered on the continuum position, at least along the slit, which
is aligned at a position angle of 5\degr, or roughly perpendicular to the disk plane. 
We tentatively identify the  emission features as
the H$_2$ S(3) line at 9.66\,$\mu$m and the out-of-plane bending mode of PAHs at 11.2\,$\mu$m. The strength of the S(3) line corresponds to an integrated flux of 
$(9\pm 2)\times 10^{-15}\rm\, erg\,s^{-1}\,cm^{-2}$, while the 11.2\,$\mu$m line flux is 
$(15\pm 3)\times 10^{-15}\rm\, erg\,s^{-1}\,cm^{-2}$, using a 1st order polynomial fit to the continuum. For an excitation temperature of $\sim 1000\,$ K 
the other H$_2$ lines are expected to be below the detection limit, taking into account the higher noise at the 6.9\,$\mu$m S(5) line.
The PAH line flux is 2-3 orders of magnitude less and the line-to-continuum ratio is about 10 times larger than those
determined for other PAH-emitting disks in the c2d sample \citep{Geers06}.

\begin{figure}
  \plotone{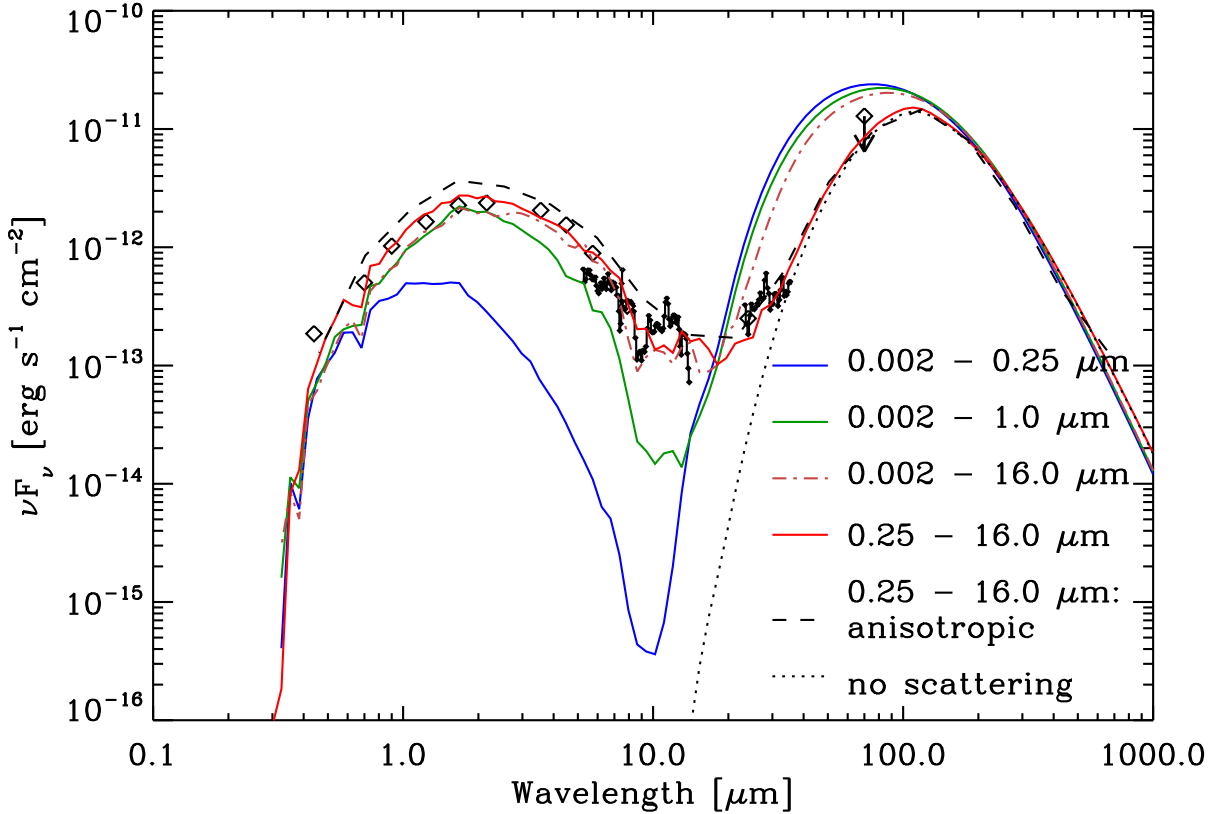}
  \caption{Model fit to the SED of the ``Flying Saucer''. The optical and near-infrared 
points are taken from \cite{Grosso03}. Grain size distributions weighted toward increasing grain sizes move the
mid-infrared minimum in the SED to progressively longer wavelengths. The dashed curve the same as model 3, but with a 
proper implementation of anisotropic scattering. The dotted curve shows the SED with the scattered photons removed. 
No photons directly from the central star are seen at any wavelength. The colors of the curves refer to the same grain size distributions 
as those in Fig. \ref{saucer_grains}. }
  \label{saucer_sed}
\end{figure} 

The spectral energy distribution (SED), shown in Fig. \ref{saucer_sed}, declines between 4 and 14\,$\mu$m. 
The LL spectrum falls below the detection limit of $\sim 1\,$mJy below 24\,$\mu$m, but rises steeply at longer wavelengths. 
This spectral shape closely resembles that theoretically 
expected for a circumstellar disk viewed close to edge-on \citep[see e.g.][]{Whitney03}.
The observed mid-infrared fluxes of 1 mJy matches closely those
predicted by the \cite{Grosso03} model. At short wavelengths ($\lambda\lesssim 40\,\mu$m) 
the optical depth to the source (the star and hot dust in the inner disk) is much greater than
unity. This means all the short-wavelength photons are those that are scattered off the surface of the disk at large radii - in this case, 50--300\,AU. 
The SED of the scattered light will have a shape that can be calculated roughly as the spectrum of the inner star+disk system
multiplied by the albedo of the dust in the outer disk.
At long wavelengths, the SED is dominated by thermal emission from the cold, outer part of the disk. These two
components create a characteristic double-peaked SED with a minimum in the range 5-20\,$\mu$m that is unrelated to, but may
in some cases be confused with, the silicate band at 9.7\,$\mu$m

\begin{table}
\centering
\caption{Mid-infrared photometry of the ``Flying Saucer''}
\begin{tabular}{lll}
\hline
\hline
Wavelength ($\mu$m)&Flux [mJy]&Reference\\
\hline
3.6&$2.43\pm 0.24$&c2d\\
4.5&$2.32\pm 0.23$&c2d\\
5.7&$1.70\pm 0.17$&c2d\\
7.9&$0.84\pm 0.08$&c2d\\
24&$2.0\pm 0.5$&c2d\\ 
70&$<300$&c2d\\
\hline
\end{tabular}
\label{photometry}
\end{table}

\section{Constraining the dust size distribution}

\begin{figure}
  \includegraphics[angle=90,width=16cm]{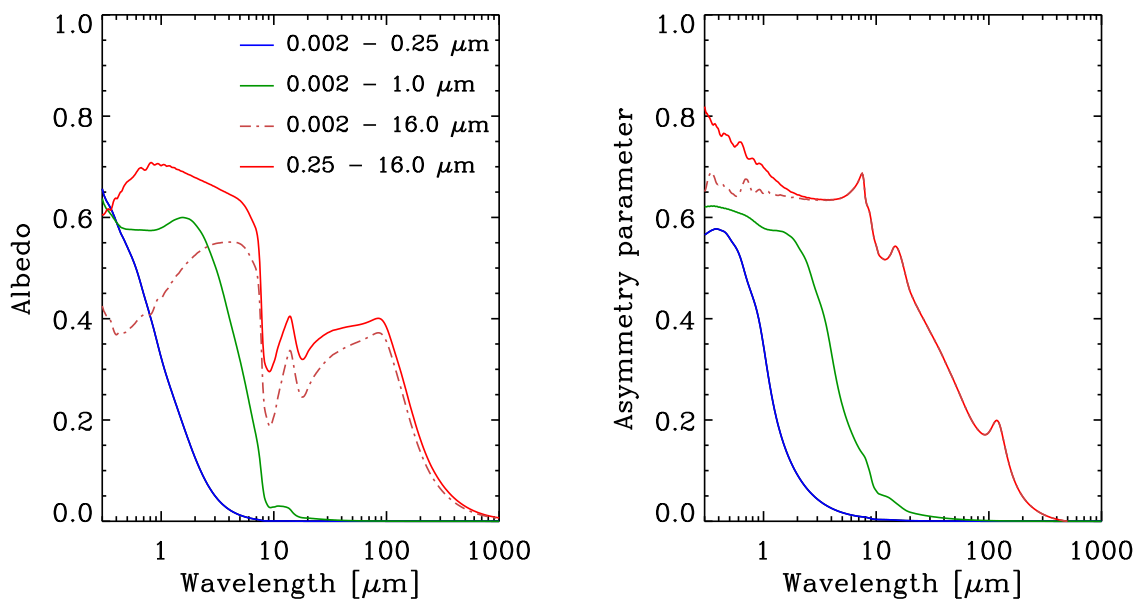}
  \caption{Albedos and the angle-averaged phase functions or ``asymmetry parameters'' (${\rm <cos\Theta>}=g$) for different grain size distributions.
In principle, the albedo is a measure of the ability of a grain population to scatter
light into the line of sight without removing it through absorption and re-emission at longer wavelengths.}
  \label{saucer_grains}
\end{figure}

We have constructed a 2D continuum radiative transfer model of the ``Flying Saucer'' with a 
setup similar to that of \cite{Pontoppidan_crbr} using the Monte Carlo code RADMC \citep{Dullemond04}. This
setup uses an axisymmetric density structure on a polar grid to calculate images and spectra.  
We use Mie theory to calculate the input 
opacities to the model and explore a range of different grain size distributions.
These are modeled by a power law size distribution $dn(a)/da\propto a^{-\alpha}$ with maximum and minimum grain radii, 
$a_{\rm min}$ and $a_{\rm max}$ and $\alpha=3.5$. The dust material is silicate with optical constants from \cite{Draine84} and
amorphous carbon (800 K experiment of \cite{Jaeger98}). For simplicity, it is assumed that carbon and silicate grains are separate populations with 15\%  of the mass in 
carbon. Fig. \ref{saucer_grains} shows the scattering properties for some of the grain models. 

We model the physical disk structure using the \cite{Grosso03} model parameters as a starting point. 
These are an outer disk scale height of $H(100\,\rm AU)/100\,AU=0.15$, a disk mass of $2\times 10^{-3}\,M_{\odot}$,
a surface density power law $\Sigma(R)\propto R^{-1}$ and a flaring scale height that varies as: $h\propto R^{5/4}$. The
disk has a radius of 300\,AU. The disk parameters have
been obtained by \cite{Grosso03} by detailed fitting to the near-infrared images, using a grain size distribution
with a maximum grain size of $\sim 50\,\mu$m. Since
no photometry is available above 70\,$\mu$m, the properties of the disk mid-plane are not well-constrained. 
Some changes to the \cite{Grosso03} model were made. First, we adopted a slightly smaller distance of 125\,pc instead of 140\,pc \citep{deGeus89}. Second, 
to improve the fit to the near-infrared and IRAC photometry, we found it necessary to include a relative 
accretion luminosity of $L_{\rm acc}/L_{*}=0.6$ to the system, 
modeled by adding a blackbody with a temperature of 1400\,K to the input 
stellar spectrum, which is modeled by a 0.084\,$L_{\odot}$ 3500 K stellar atmosphere from \cite{Kurucz79}, as well as setting the
foreground extinction to $A_V=0.5$\,mag. The accretion luminosity, $A_V$ and effective temperature are somewhat interdependent, so 
other combinations may be possible, in particular, the central star may be cooler and more luminous, requiring less accretion. 
Removing the accretion component entirely will lower the
1-5\,$\mu$m flux by a factor of $\sim 2$. 
In any case, these parameters do not affect wavelengths longer than $\sim 5\,\mu$m, and therefore not the derived grain size distribution. 

\begin{figure}
  \plotone{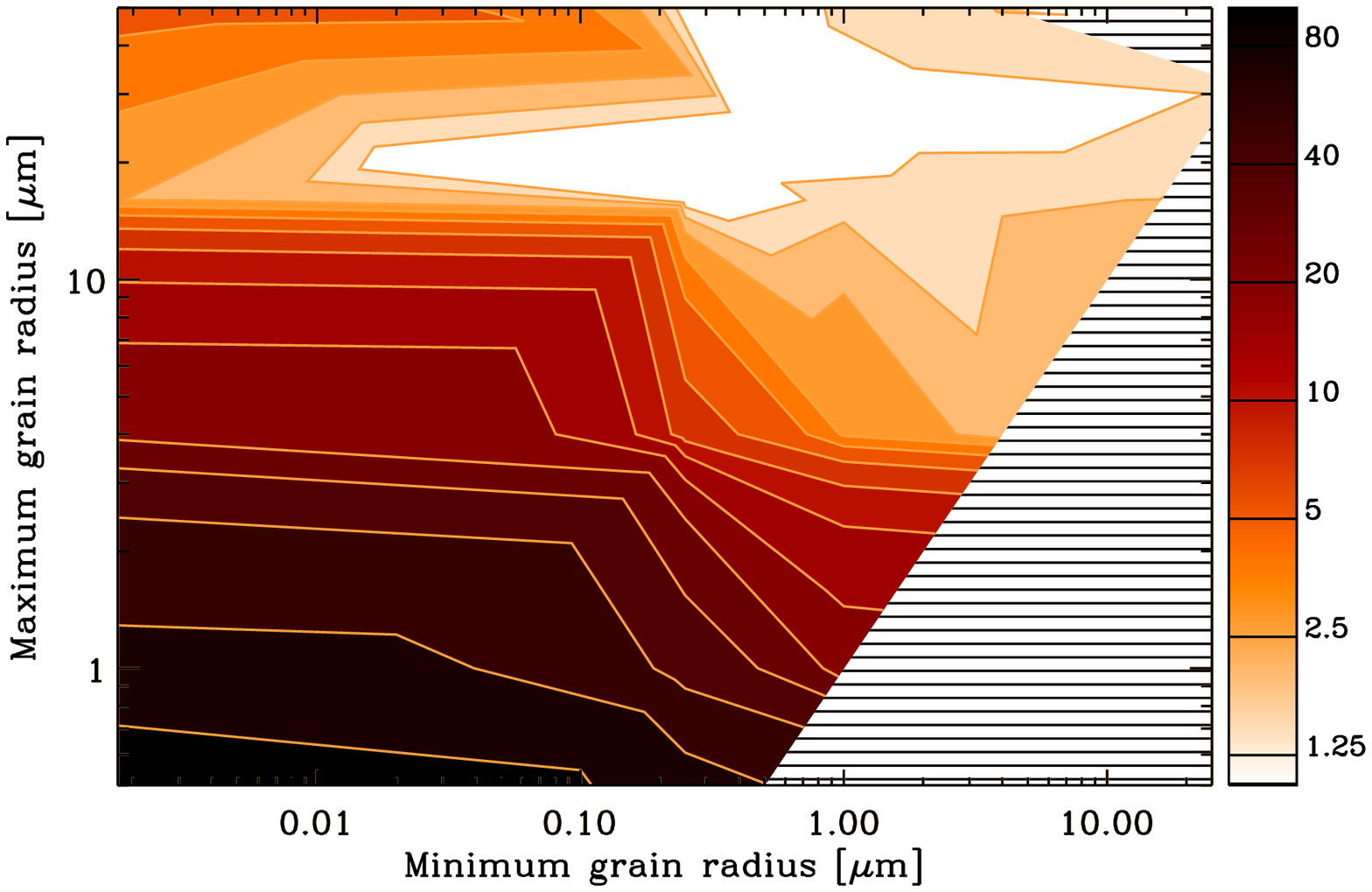}
  \caption{Goodness-of-fit surface for the calculated grid of models. The contours have been normalized, such that 
the best-fitting model has a goodness-of-fit of unity. The contours shown are in steps of 40\%, the lowest having a value 
of 1.4. The hashed pattern indicates the area where no reasonable models can been calculated because 
$a_{\rm min}>a_{\rm max}$ . }
  \label{chi2}
\end{figure}

Given a set of parameters determining the physical structure of the disk the grain size distribution
strongly affects the shape and the location of the minimum of the SED in the mid-infrared. In Fig. \ref{saucer_sed}, 
models with different grain size distributions are compared to the observed SED and in particular to the IRS spectrum.
As larger grains are included in the grain population, the mid-infrared flux due to scattered photons increases relative
to the rest of the SED. While light is always scattered at the surface where the scattering optical depth ($\tau_{\rm sca}$) is unity, the albedo 
controls how much light actually survives (is not absorbed and re-emitted at longer wavelengths) 
to be scattered at $\tau_{\rm sca}=1$. According to Fig. 
\ref{saucer_grains}, a significant population of $\sim 10\,\mu$m grains is required in order to have an 
albedo\,$=Q_{\rm sca}/(Q_{\rm abs}+Q_{\rm sca})\gtrsim 0.1$ at 20\,$\mu$m.  
Additionally, as the small grains are removed, the thermal emission peak shifts to longer wavelengths.

We ran a grid of models varying the minimum and maximum grain sizes, while keeping the density structure constant, to minimize the quantity 
$\int (F_{\nu,\rm model}(\nu)-F_{\nu,\rm obs}(\nu))^2d\nu$. Figure \ref{chi2} shows the
goodness-of-fit surface. Inspection shows that good matches to the SED are 
found for grain size distributions with $a_{\rm min}\gtrsim 0.5\,\mu$m and $a_{\rm max}\gtrsim 10\,\mu$m. 
While the exact limits to the grain size distribution are not strictly constrained from the SED alone, 
the presented model clearly excludes grain size distributions
with no grains larger than $\sim 5\mu$m or distributions that are dominated (in mass) by grains that are $\lesssim 0.5\,\mu$m.
The upper cutoff for the grain size distribution may be significantly larger. The required presence of large grains is very stable
to changes in the density structure of the disk and slope of the grain size distribution. Whether there is a lower cutoff to the grain size is less secure. For instance, 
a shallower grain size distribution was found to, at least in part, lessen the need for a lower cutoff of grain sizes. Additionally, a larger disk
mass may be required since our best-fitting J-band opacity is $\sim 3$ times lower than that used by \cite{Grosso03}. Models with higher disk masses
lessen the need for a lower grain size cutoff by shifting the thermal peak in the SED to longer wavelengths, but do not affect the scattering peak. 
We are therefore confident that large grains are present in the Flying Saucer, while the evidence for a lack of small grains is somewhat more ambiguous.  
This result strengthens similar conclusions obtained for other edge-on disks using less constraining data at shorter wavelengths 
\citep{Wood02, McCabe03} and provides strong evidence for the presence of grain growth in the outer disk of the
``Flying Saucer''. 

The model grid uses isotropic scattering, thereby requiring much fewer photon packages than if 
anisotropic scattering is used. While not a proper treatment of scattering of large grains, computing time is drastically cut this way.
To ensure that the conclusions are not affected, we also calculated the SED of the best-fitting model
using a proper anisotropic scattering model. Anisotropic scattering has the effect of making the image of the disk more compact around the
central star. The effect on the SED is small as seen in in Fig. \ref{saucer_sed}; it
changes the flux by $\sim 50\%$ at wavelengths dominated by scattered photons. 
This difference is very small compared to the 2-3 orders of magnitude dependence on the grain size distribution in the mid-infrared.

\section{Discussion}

The IRS spectrum of the ``Flying Saucer'' has provided important constraints on the nature of disk material in the surface layers ($H/R\sim 0.1$) of the disk
at radii {\it larger} than 50\,AU in several different ways. First, dust grains in this part of the disk are much larger than grains
in the interstellar medium, clearly implying significant grain growth. Second, tentative emission features from PAHs and H$_2$ reveal clues about 
emission excited by UV photon
processes in the outer disk. Had the disk been viewed face-on, these faint features would have disappeared in the glare from the warm inner parts of the disk;
at 10\,$\mu$m the ``Flying Saucer'' would be 200 times brighter if viewed face-on.

Evidence from silicate emission features has already indicated that grains have grown to several $\mu$m \citep{Bouwman01, Boekel05}
{\it at small disk radii}, within $\sim 10\,$AU \citep{Boekel04}, in Herbig Ae disks, as well as T Tauri stars \citep{Kessler-Silacci06}. 
In contrast, mid-infrared observations of scattered light such as those presented in this letter
can reveal the presence of large grains in the surface layers much further away from the central star. While millimeter continuum studies have
also found evidence for large grains in the outer regions of disks \citep[e.g.][]{Natta04, Rodmann06}, such observations probe only the mid-plane. 
In the case of the ``Flying Saucer'', we have shown that significant grain growth has taken place at large distances from the star (up to several 100\,AU), consistent
with the conclusions of \cite{McCabe03} in the case of HK Tau B, a similar edge-on disk. 
When comparing this result to recent models of grain growth and settling \citep[e.g.][]{Dullemond04, Dullemond05}, an apparent problem appears. These
models predict that grain growth to sizes $> 10\,\mu$m followed by rapid settling to the disk mid-plane takes place on very short timescales ($10^3-10^5$ years).
This will effectively remove larger grains from the disk surface on a time scale that increases with radius and decreases with grain size. 
The new observational evidence presented here shows that large grains are present in the disk surface 
not only close to the star, as found in other studies, but also at much larger distances. This implies that in both these regimes there is an efficient mechanism
for preventing grains in the $\sim 10\,\mu$m range from settling. 

Adding to the puzzle is the indication of the presence of very small grains, such as PAHs, in the same
general region of the disk as the large grains. Even if the detection of PAHs in the ``Flying Saucer'' is only tentative, 
some edge-on disks clearly show very strong PAH features \citep{Perrin06}. Very small grains will not settle, so
the question they pose is why they have not been removed by the grain coagulation producing the observed population of large grains, or whether there is a 
process to replenish them.  

Another possibility for explaining the presence
of PAH emission is that the 11.2$\,\mu$m feature may be light from the inner parts of the disk scattered on the 
outer surface. This requires that the line-to-continuum ratio of the PAH emission from the inner disk is the same as that observed in the ``Flying Saucer''. 
Observations of face-on disks have shown that such strong line emission is rare, but not impossible \citep{Geers06}.

The tentative detection of the S(3) line of H$_2$ is interesting. The S(3) line
dominates the H$_2$ line spectrum for gas at a temperature of a few hundred Kelvin. 
Detailed modeling by \cite{Nomura05} has shown that, in the presence of a significant
UV excess, the S(3) line flux from a disk very similar to that of the ``Flying Saucer''
is expected to be $\sim 9\times 10^{-15}\,\rm erg\,s^{-1}\,cm^{-2}$, with most of the flux
originating in the uppermost layers of the disk at radii of 10-100\,AU. If confirmed, the presence
of the line can therefore be interpreted as an indicator of a strongly enhanced UV field in the upper
layers.

Finally, the possible presence of high contrast emission lines in the spectrum of the ``Flying Saucer'' suggests
that edge-on disks may be excellent targets for looking for fluorescent tracers in the outer disk that 
would not have been seen if the disk had been viewed face-on. 
This illustrates the need for sensitive space-born high-resolution mid-infrared spectroscopy in the future offered by
SOFIA and the James Webb Space Telescope.

\acknowledgements{
We are grateful for a constructive review by an anonymous referee.
KMP is supported by NASA through Hubble Fellowship grant \#01201.01 awarded by the Space Telescope Science Institute, 
which is operated by the Association of Universities for Research in Astronomy, Inc., for NASA, under contract NAS 5-26555.
Support for this work, part of the Spitzer Space 
Telescope Legacy Science Program, was provided by NASA through Contract Numbers 1224608 and 
1230779 issued by the Jet Propulsion Laboratory, California Institute of Technology under NASA contract 1407. 
EvD acknowledges an NWO Spinoza prize. 
}

\bibliographystyle{apj}
\bibliography{ms}

\end{document}